\documentclass{vgtc}                          
\ifpdf
  \pdfoutput=1\relax                   
  \usepackage{graphicx}                
  \DeclareGraphicsExtensions{.pdf,.png,.jpg,.jpeg} 
\else
  \usepackage{graphicx}                
  \DeclareGraphicsExtensions{.eps}     
\fi%

\graphicspath{{figures/}{pictures/}{images/}{./}} 

\PassOptionsToPackage{warn}{textcomp}  
\usepackage{textcomp}                  
\usepackage{mathptmx}                  
\usepackage{times}                     
\usepackage{cite}                      
\usepackage{tabu}                      
\usepackage{booktabs}                  

\usepackage[normalem]{ulem}
\usepackage{xcolor}

\newcommand{\removed}[1]{}

\usepackage{multirow}

\vgtccategory{Research}

\title{Virtual Steps: The Experience of Walking for a Lifelong Wheelchair User in Virtual Reality}

\newcommand{\equalcontribution}{\thanks{These authors contributed equally to this work.}}

\author{Atieh Taheri*\equalcontribution\thanks{e-mail: a\_taheri@ucsb.edu} %
\and Arthur Caetano*\thanks{e-mail: caetano@ucsb.edu} %
\and Misha Sra\thanks{e-mail: sra@ucsb.edu}}
\affiliation{\scriptsize University of California, Santa Barbara}


\vgtcauthor{Atieh Taheri, Arthur Caetano, Misha Sra}

\vgtcinsertpkg

\abstract{
Many people often take walking for granted, but for individuals with mobility disabilities, this seemingly simple act can feel out of reach. This reality can foster a sense of disconnect from the world since walking is a fundamental way in which people interact with each other and the environment. Advances in virtual reality and its immersive capabilities have made it possible to enable those who have never walked in their life to ``virtually'' experience walking. We co-designed a VR walking experience with a person with Spinal Muscular Atrophy who has been a lifelong wheelchair user. Over 9 days, we collected data on this person's experience through a diary study and analyzed this data to better understand the design elements required. Given that they had only ever seen others walking and had not experienced it first-hand, determining which design parameters must be considered in order to match the virtual experience to their idea of walking was challenging. Generally, we found the experience of walking to be quite positive, providing a perspective from a higher vantage point than what was available in a wheelchair. Our findings provide insights into the emotional complexities and evolving sense of agency accompanying virtual walking. These findings have implications for designing more inclusive and emotionally engaging virtual reality experiences.

} 

\CCScatlist{
  \CCScatTwelve{Human-centered computing}{Accessibility}{Empirical studies in accessibility}{};
  \CCScatTwelve{Human-centered computing}{Human computer interaction (HCI)}{Interaction paradigms}{Virtual reality}
}

\begin{document}

\maketitle

\section{Introduction} 

Walking is more than a mere means of locomotion; it is an action that enables freedom, exploration, and independence. Certain conditions and disorders can prevent individuals from walking, including genetic disorders and congenital conditions such as Cerebral Palsy, Muscular Dystrophy, or Spinal Muscular Atrophy. These conditions can affect the development of muscles, bones, or the nervous system from birth and hinder the ability to walk. 

Walking produces a range of sensory stimuli: proprioceptive feedback generated by limb movement~\cite{goble2006upper}, vestibular feedback from head motion~\cite{roy2001selective}, optical flow from visual perception as they move~\cite{bruggeman2007optic}, and auditory feedback produced by their footsteps~\cite{turchet2013walking}. Emulating these sensations is a complex task that has been the target of extensive research, especially using virtual reality. Notable methods include employing specialized equipment such as omnidirectional treadmills~\cite{darken1997omni, iwata1999walking}, foot-support motion platforms~\cite{iwata2001gait}, and walking spheres~\cite{medina2008virtusphere}. VR technology has been used to simulate walking, providing presence and embodiment~\cite{moura2021embodiment, riva2019neuroscience, kim2017walking}.

However, previously proposed VR walking experiences were mainly focused on reproducing the experience of a person who is able, or at some point in the past, was able to walk. This focus is not necessarily aligned with the desires of those users who have never experienced standing and walking. Even though these users have never walked, they have developed a mental model of how it feels to walk by observing others walk and interacting with them. Due to the potential differences in these mental models, current systems may not respond in ways that users who have never walked anticipate~\cite{norman2013design}. To address this gap, it is crucial to develop VR walking simulations that specifically cater to the unique mental models of individuals who have never experienced walking, ensuring a more inclusive and effective user experience. Another limitation of prior VR walking experiences is that they rarely consider the interaction limitations that usually accompany mobility limitations. Users who never walked before may also face challenges using the input devices and interaction techniques offered in commercial VR devices because they rely on fine hand and head coordination. The lack of appropriate interaction design restricts the access of individuals with congenital motor and mobility impairments to walking experiences in VR.

In this study, we interactively co-designed a VR walking experience with a person with Spinal Muscular Atrophy who is a lifelong wheelchair user, delving into their perceptions and emotional responses to this novel sensation. We explored different design alternatives including auditory feedback, visual representation, movement dynamics, walking speed, interaction modality, and virtual environment. Instead of concentrating on rehabilitation or therapeutic uses, our research explores the design parameters of a VR walking experience tailored to users who have no prior physical walking experience. The needs of this particular audience can be quite distinct from those of the general population because they lack a physical walking experience to use as a reference, except for mobility aided by devices like wheelchairs. Our contributions are as follows:
\begin{enumerate}
    \item An initial set of features and parameter values for VR walking applications to individuals with congenital motor and mobility impairments.
    \item Qualitative insights into the emotions and perceptions of a user with congenital motor and mobility impairments while experiencing walking through VR.
\end{enumerate}

Through our study, we found an initial set of factors that affect walking experiences in VR for users who have never walked and compiled how different strategies of walking sensation emulation can impact the user experience. This contribution can inform the design of next-generation VR experiments that target audiences with motor and mobility disabilities, broadening accessibility in VR and providing a novel experience to users who never walked in their lives, with potential psychological and cognitive impacts.

\section{Related Work}

\subsection{Simulating Walking in VR}
The use of VR to simulate novel experiences has been an active area of research that spans multiple disciplines, including neurological rehabilitation, where VR techniques resulted in both physical and cognitive enhancements for mobility and walking~\cite{holden2005virtual, voinescu2021virtual}. Many studies have harnessed VR technology to simulate walking environments, which is crucial for gait rehabilitation~\cite{janeh2021review}. Typically, gait rehabilitation involves walking either overground or on treadmills~\cite{park2013comparison}. Moan et al.~\cite{moan2021experiences} developed a fully immersive virtual reality treadmill game for post-stroke rehabilitation. They found that using VR games as part of gait rehabilitation after stroke is acceptable and potentially useful, based on positive experiences from testing by both stroke survivors and clinicians. Differently, our work focuses on reproducing the visual and auditory sensations of walking in VR experience for individuals who have never walked. We aim to emulate these sensations while preserving the user's agency.

There are limitations in simulating realistic walking sensations for those who have never walked before. The study by Matsuda et al.~\cite{matsuda2021enhancing} proposed a new virtual walking system that combines optic flow, foot vibrations, and a walking avatar to induce a sense of virtual walking in seated users without limb motion. In a different method, the idea of ``King-Kong Effects'' was presented to augment the feeling of locomotion in virtual settings~\cite{terziman2012king}. Drawing motivation from cinematic special effects, the researchers suggested incorporating either visual or tactile vibrations with every emulated footstep while navigating virtually. Saint-Aubert et al.~\cite{saint2022influence} investigated the impact of user posture and virtual exercise on locomotion perception in VR. Participants, wearing a VR headset, watched a first-person avatar doing virtual exercises, revealing that the sensation of walking could be adapted to various postures and exercises, indicating avenues for enriching the virtual walking experience in diverse settings.

In these studies, participants did not have mobility impairments and had their prior walking experience as a baseline to compare against the virtual experience. Our work pursues a different goal, providing walking sensations for individuals who have never walked before and do not have prior experience to anchor their mental model of how walking sensations feel. Hence, investigating the factors and principles in VR that could either enhance or hinder their experiences is crucial for the satisfactory design of VR applications for this audience.

\subsection{Psychological Aspects of Locomotion in VR}
\subsubsection{Presence and Embodiment in VR Locomotion}

Presence, often described as the feeling of \emph{being there} in a virtual environment, is amplified when users can simulate actions intrinsic to their daily lives, such as walking~\cite{wienrich2021spatial, uz2022processing, hruby2020empirical, engel2008psychophysically}. North and North\cite{north2016comparative} suggest in their study that the sensation of presence in VR is directly influenced by the user's ability to interact with the environment naturally and intuitively. In the context of walking, this means the more realistic the simulation of locomotion, the greater the user's sense of presence. The sense of presence is particularly evident when users navigate vast virtual terrains or intricate pathways, feeling genuinely immersed in the environment~\cite{yoon2006novel}. 
Habgood et al.~\cite{habgood2018rapid} found that teleportation, a commonly used locomotion technique, reduces spatial presence compared to other locomotion techniques such as free movement and node-based navigation. In a VR power wheelchair simulator, researchers have found three factors that affect the sense of presence: display type, control over the field of view, and visualization of the user's avatar~\cite{alshaer2017immersion}.

Embodiment in virtual reality goes a step further, focusing on the user's connection with their virtual avatar or representation~\cite{matamala2019immersive}. More than simply displaying a virtual representation of the user; embodiment happens when there is a feeling of genuine connection to a virtual body. When users see that their virtual legs moving in sync with their intended actions, it bridges the gap between their real and virtual selves.
For example, Banakou and Slater~\cite{banakou2014body} conducted experiments that demonstrated how multisensory stimulation can induce a sense of virtual body ownership. Their findings suggest that when visual input (seeing a virtual body) is combined with synchronous tactile input (feeling touch), participants can experience the illusion of owning a virtual body. This research underscores the importance of multi-sensory integration in achieving a strong sense of embodiment in VR. Additionally, Okumura et al.~\cite{okumura2020investigating} highlighted the role of proprioceptive drift in enhancing the sense of embodiment. They found that when there's a spatial congruence between the real and virtual bodies, users tend to perceive their actual body position closer to the virtual one, further strengthening the sense of embodiment.

In summary, the sensation of presence and embodiment in VR, especially in the context of locomotion, is influenced by various factors, including the realism of the virtual environment, the user's ability to interact naturally, the first-person visualization of an avatar, and the congruence between visual and physical cues. These prior findings have informed our experimental design that allows users to visualize themselves as a first-person avatar in an HMD VR device that plays a walking animation when the user is moving. It facilitates control of the direction and speed of a continuous movement and induces a head oscillation that emulates the visual experience of walking, and audio feedback of footsteps. The HMD position is tracked and used to move the avatar head. The user can also visualize the avatar through the reflection of a virtual mirror.

\subsubsection{Agency and Control in VR Locomotion}
Agency refers to the capacity for a user to make decisions and take meaningful actions that have effects within the virtual world. A systematic review by Radianti et al.~\cite{radianti2020systematic} underscores two clear tiers of user engagements: 1) those taking place within the VR realm, and 2) those entailing hardware interactions, illuminating a structure of agency fostered by VR technology. The interplay of agency and control in locomotion in VR is particularly essential to enhance user experience. Cardoso et al.~\cite{cardoso2019survey} examined real locomotion methods and highlighted the importance of interaction strategies, amalgamating devices, user engagements, and system reactions, to nurture agency and control within VR settings. Schafer et al.~\cite{schafer2022controlling} investigate continuous locomotion control through hand gestures, providing innovative methods to improve user control in VR locomotion endeavors, thereby promoting user agency~\cite{schafer2022controlling}.

All prior research revolved around the sense of agency when walking or in motion in VR for people who had previously experienced walking in the real world. However, it is crucial to also consider those who have no prior real-world walking experience and their mobility modality was different and investigate how their sense of agency is affected.

The Bayesian integration model of Sense of Agency states human judgment of agency is influenced not only by the sensorimotor systems but also by factors at the cognitive level~\cite{wen2022sense, legaspi2019bayesian, moore2012sense}. According to this model, both cognitive and sensorimotor signals go through Bayesian integration where the lower their variance the higher their weight. Based on this model we expect a user with a congenital sensorimotor impairment to still experience agency in a virtual walking environment thanks to a mental model of the phenomena that occurs at the cognitive level.

\subsection{Cybersickness}

Cybersicnkess is a common problem in VR, often attributed to the sensory mismatch between visual and vestibular systems that occurs when users receive visual cues of motion without corresponding vestibular stimulation~\cite{davis2014systematic}. A study showed that $68\%$ of the users have experienced cybersickness in a virtual flight session~\cite{cevette2012oculo}. This syndrome is also more common among female individuals~\cite{munafo2017virtual}. Prior work has investigated techniques to reduce cybersickness including the use of galvinic vestibular stimulation~\cite{cevette2012oculo, sra2019adding}. In an evaluation of a wheelchair simulator, users of power wheelchairs also reported moderate cybersickness~\cite{vailland2021vr}, suggesting that this audience is also affected by cybersickness regardless of a possible similar sensory conflict during the use of assistive devices for locomotion. In our work, we carefully adjusted speed parameters to mitigate discomfort and cybersickness. The system initially implemented a snap-turn technique, but in later versions, we adopted a continuous angular movement at a comfortable speed for the user. The head oscillation parameters were also tuned carefully.

\section{Method}

To investigate the design requirements for emulating walking sensation for individuals who have never walked, we applied participatory design coupled with a \textit{diary study} technique, i.e., one of the co-authors, Taheri is a member of the target audience and was the single participant of the study, keeping observations derived from the use of the experimental system in a diary. The participant had an active role not only in testing the system, but also in guiding, requesting features, and calibrating parameters to meet their needs. Diary studies have been successfully employed in disability research, leading to novel findings on the experience of people with disabilities in VR experiences~\cite{zhang2023diary}.
Our decision to use a diary study over phenomenological auto-ethnography was guided by our objective to allow for the potential generalizability of our findings. While phenomenological auto-ethnography offers deep, introspective insights into a researcher’s personal experience, it focuses on subjectivity and is less concerned with generalizability~\cite{grant2015demedicalising}. A relevant example of a similar approach in VR research for people with disabilities is the diary study conducted by Zhang et al.~\cite{zhang2023diary} to explore the impact of avatars with disability signifiers on the experiences of people with disabilities in social VR settings. We also observed the previously proposed recommendations for HCI research with autobiographical design~\cite{desjardins2018revealing}. Our participant and co-author is identified as a member of the community of people who had never walked in their lives. According to Liang et al.~\cite{liang2021embracing}, membership implies shared or similar goals and experiences of an individual with those in a group and can benefit research in multiple ways, including supporting expertise in making experiment choices and interpreting results from an insider perspective.

\subsection{Participant}
Taheri, a 35-year-old female and the co-author of this paper served as the sole participant in the study. Taheri was born and has been living with Spinal Muscular Atrophy (SMA) type 2, a severe genetic motor neuron disease. Due to the nature of SMA type 2, she has never had the ability to stand up, let alone walk. Consequently, she relies on a wheelchair for mobility. She possesses both a mechanical wheelchair and a power wheelchair, which is controlled via a joystick. However, it is important to note that Taheri requires assistance to place her hand on the joystick and can only move her right thumb.

While Taheri has some familiarity with VR development, her hands-on experience has been limited due to the inaccessibility of VR controllers. This background provides a distinct perspective on VR interaction, particularly in terms of accessibility challenges.

The decision to have the co-author as the participant was intentional. By focusing on someone who has never had the ability to stand or walk, the study aims to provide a deeply personal perspective on the experiences and emotions evoked by virtual walking scenarios. This unique partnership between the researcher and the participant enhances the richness and depth of the insights derived from the research. Throughout the study, Taheri maintained a diary to record her feelings, perceptions, and reflections after each VR session. This diary became a primary data source for sentiment analysis, capturing her evolving experiences and reactions to the varying virtual walking scenarios. Given that Taheri is also a co-author, she played an active role in shaping the research objectives, providing feedback on the virtual environment and adjustable parameters, and reviewing preliminary findings. Her dual role provided a valuable iterative feedback loop throughout the research process, ensuring both accuracy and authenticity in interpreting the experiences.

\subsection{Procedure}

This experiment was conducted with the informed consent of the participant and was approved by the local IRB of our home institution. We adopted a participatory design approach in $9$ iterations. Starting from a baseline implementation of the experimental system, Taheri tested the system for a total duration of 3 hours and 45 minutes. In line with the principles of participatory design, we did not impose a fixed duration for each trial session~\cite{schuler1993participatory}. Instead, Taheri was given the freedom to engage with the VR system as long as desired during each session, allowing for a natural and in-depth exploration of the virtual space and the available features. This approach enabled Taheri to thoroughly engage with the VR system and reflect on her experience in a manner that was not constrained by time limits. Figure~\ref{fig:Fey-mirror}:a depicts Taheri engaged in one of her VR explorations within the experimental system. 

Following each VR session, Taheri proceeded to take notes in her diary, checking for changes compared to the previous version, and constantly evaluating certain aspects like functionality and user experience across all iterations. These diary entries were made immediately after each session, using her right thumb on her laptop's built-in touchpad and a virtual keyboard, as is her usual method. This approach was chosen to capture her immediate reactions and thoughts more authentically. After the trial and journaling, Taheri and Caetano, the other co-author, who worked as the developer, discussed the challenges faced by Taheri and idealized possible solutions based on the feedback provided. These were then implemented in the following version by Caetano. Using this iterative approach allowed Taheri to thoroughly evaluate the system while it was being designed and provide targeted feedback for improvements.

\subsection{Experimental System}

\begin{table*}[t!]
\centering
\begin{tabular}{|>{\centering\hspace{0pt}}m{0.1\linewidth}|>{\hspace{0pt}}m{0.84\linewidth}|} 
\hline
\textbf{Iteration} & \multicolumn{1}{>{\centering\arraybackslash\hspace{0pt}}m{0.84\linewidth}|}{\textbf{Modifications \& Feedback}} \\ 
\hline
1 & Introduced basic VR walking simulation. Feedback: Encountered issues with recognition of ``left'' and ``back'' commands. Voice recognition improvements are required. Faced an issue with backward movement. \\ 
\hline
2 & Changed ``walk'' command to ``go''; introduced head oscillation for movement realism. Fixed previous issues with backward movement. Feedback: Adjustment for head oscillation required. A bug in terrain navigation required fixing. \\ 
\hline
3 & Included avatar and footsteps sound; added options to toggle avatar, head oscillation, and footsteps; fixed the bug with gravity; introduced log file for tracking. Feedback: Encountered issues with the avatar's head position, footsteps sound being out of sync with head oscillation, and slow walking speed. Head oscillation felt more unnatural and needed correction. Improvement is needed for avatar integration and movement speed. \\ 
\hline
4 & Adjusted head oscillation and walking speed; better synchronization of footsteps and head movements; fixed head position bug. Feedback: Persistent issues with ``left'' and ``back'' commands. Further refined voice commands are required. \\ 
\hline
5 & Implemented ``lima'' for left turns. Feedback: Felt monotony during long-distance virtual walks. Required having control over movement speed and direction. \\ 
\hline
6 & Implemented ``reverse'' for moving back; added speed control commands; allowed adjustment of oscillation amplitude. Feedback: Experienced issues with ``fast'' command. Refining speed control and turning mechanism required. \\ 
\hline
7 & Replaced continuous turning with snap turning with a velocity of 1.5 degrees/sec; replaced ``fast'' with ``speed'' for acceleration. Feedback: Not satisfied with the ``speed'' command. Discussed having a ``go'' command for both walking initiation and acceleration. \\ 
\hline
8 & Implemented ``go'' for both walking initiation and acceleration. Feedback: Fully satisfied with voice commands. \\ 
\hline
9 & Final adjustments. Feedback: Celebrated improved experience and reflected on the journey. \\
\hline
\end{tabular}
\caption{A summary of modifications on the system over iterations and the corresponding feedback from Taheri.}
\label{tab:changes-feedback}
\end{table*}

\begin{table*}[ht!]
\centering
\begin{tabular}{|c|c|c|c|c|c|c|c|}
\hline
\multirow{3}{*}{\textbf{Iteration}} & \multicolumn{3}{c|}{\multirow{3}{*}{\begin{tabular}[c]{@{}c@{}}\textbf{Voice Commands}\end{tabular}}} & \multirow{3}{*}{\begin{tabular}[c]{@{}c@{}}\textbf{Base Walking}\\\textbf{Speed (meter/sec)}\end{tabular}} & \multicolumn{3}{c|}{\multirow{3}{*}{\begin{tabular}[c]{@{}c@{}}\textbf{Head Oscillation}\end{tabular}}} \\
 & \multicolumn{3}{c|}{} &  & \multicolumn{3}{c|}{} \\ \cline{2-4} \cline{6-8}
 & \textbf{Movement} & \textbf{Turning} & \textbf{Speed} &  & \begin{tabular}[c]{@{}c@{}}\textbf{Amplitude} \\ \textbf{(meter)}\end{tabular} & \begin{tabular}[c]{@{}c@{}}\textbf{Stride Len}\\\textbf{(meter)}\end{tabular} & \begin{tabular}[c]{@{}c@{}}\textbf{Adv Speed}\\\textbf{(meters/sec)}\end{tabular} \\ \hline
1 & ``walk'' + ``back'' & ``right'' + ``left'' & N/I & 0.5 & N/I & N/I & N/I \\ 
 & ``stop'' & & & & & & \\
\hline
2 & ``go'' + ``back'' & ``right'' + ``left'' & N/I & 0.5 & (0.1, 0.1, 0.1) & 0.8 & 1 \\ 
 & ``stop'' & & & & & & \\\hline
3 & ``go'' + ``back'' & ``right'' + ``left'' & N/I & 0.5 & (0.1, 0.05, 0.1) & 0.5 & 0.5 \\
 & ``stop'' & & & & & & \\ \hline
4 & ``go'' + ``back'' & ``right'' + ``left'' & N/I & 1 & (0.1, 0.1, 0.1) & 0.8 & 1 \\ 
 & ``stop'' & & & & & & \\ \hline
5 & ``go'' + ``back'' & ``right'' + ``lima'' & N/I & 1 & (0.1, 0.1, 0.1) & 0.8 & 1 \\
 & ``stop'' & & & & & & \\ \hline
6 & ``go'' + ``reverse'' & ``right'' + ``lima'' & ``fast'' + ``slow'' & 1 & (0.05, 0.05, 0.05) & 0.8 & 1 \\ 
 & ``stop'' & & & & & & \\ \hline
7 & ``go'' + ``reverse'' & ``right'' + ``lima'' & ``speed'' + ``slow'' & 1 & (0.05, 0.05, 0.05) & 0.8 & 1 \\ 
 & ``stop'' & & & & & & \\ \hline
8 & ``go'' + ``reverse''& ``right'' + ``lima'' & ``go'' + ``slow'' & 1 & (0.05, 0.05, 0.05) & 0.8 & 1 \\
 & ``stop'' & & & & & & \\ \hline
9 & ``go'' + ``reverse'' & ``right'' + ``lima'' & ``go'' + ``slow'' & 1 & (0.05, 0.05, 0.05) & 0.8 & 1 \\ 
 & ``stop'' & & & & & & \\ \hline
\end{tabular}
\caption{Changes in parameters over iterations. Voice commands include Movement, Turning, and Speed Commands. The head oscillation parameters encompass Amplitude, Stride Length (Stride Len), and Advance Speed (Adv Speed). Changes in base walking speed are also reported. "N/I" denotes "Not Implemented".}
\label{tab:params}
\end{table*}

To test our hypotheses, we developed a walking VR experience, addressing Taheri’s needs and preferences, and utilized the Unity 3D XR Interaction Toolkit\footnote{https://docs.unity3d.com/Packages/com.unity.xr.interaction.toolkit@2.3/manual/index.html}. The application was deployed to an Oculus Quest 1\footnote{https://www.meta.com/quest/}. Our system implemented a speech interface to accommodate the motor needs of Taheri. We employed Wit.ai\footnote{https://wit.ai/}, a robust speech recognition service from Meta, to transcribe and interpret user voice commands. Over the iterations, we realized that the speech recognition system was incorrectly identifying certain words as others. To address this issue, we performed some research and found that, on some occasions, using keywords from the NATO (North Atlantic Treaty Organization) phonetic alphabet might better work to improve clarity and more accurate input speech recognition. For instance, instead of using the word ``left,'' we adopted the word ``lima''.

\begin{figure}[!t]
    \centering
    \includegraphics[width=\linewidth,alt={A Unity-based system diagram depicting GameObjects on the left and Components on the right side. The elements are connected by 3 different types of relationship: (1) GameObject hierarchy, (2) Component attachment to a GameObject, and (3) References held by Components to other Components or GameObjects. "XROrigin" is at the top of the GameObject hierarchy, with "CameraOffset" directly beneath it. "CameraOffset" then parents "FootstepSFX", "Avatar", and "MainCamera". The "SpeedCharacterController" and "FeatureToggle" components are attached to the "XROrigin" GameObject. The “HeadOscillator” is attached to the main camera. The "SpeedCharacterController" component has a reference to the "AppVoiceExperience" which is a third-party component.}]{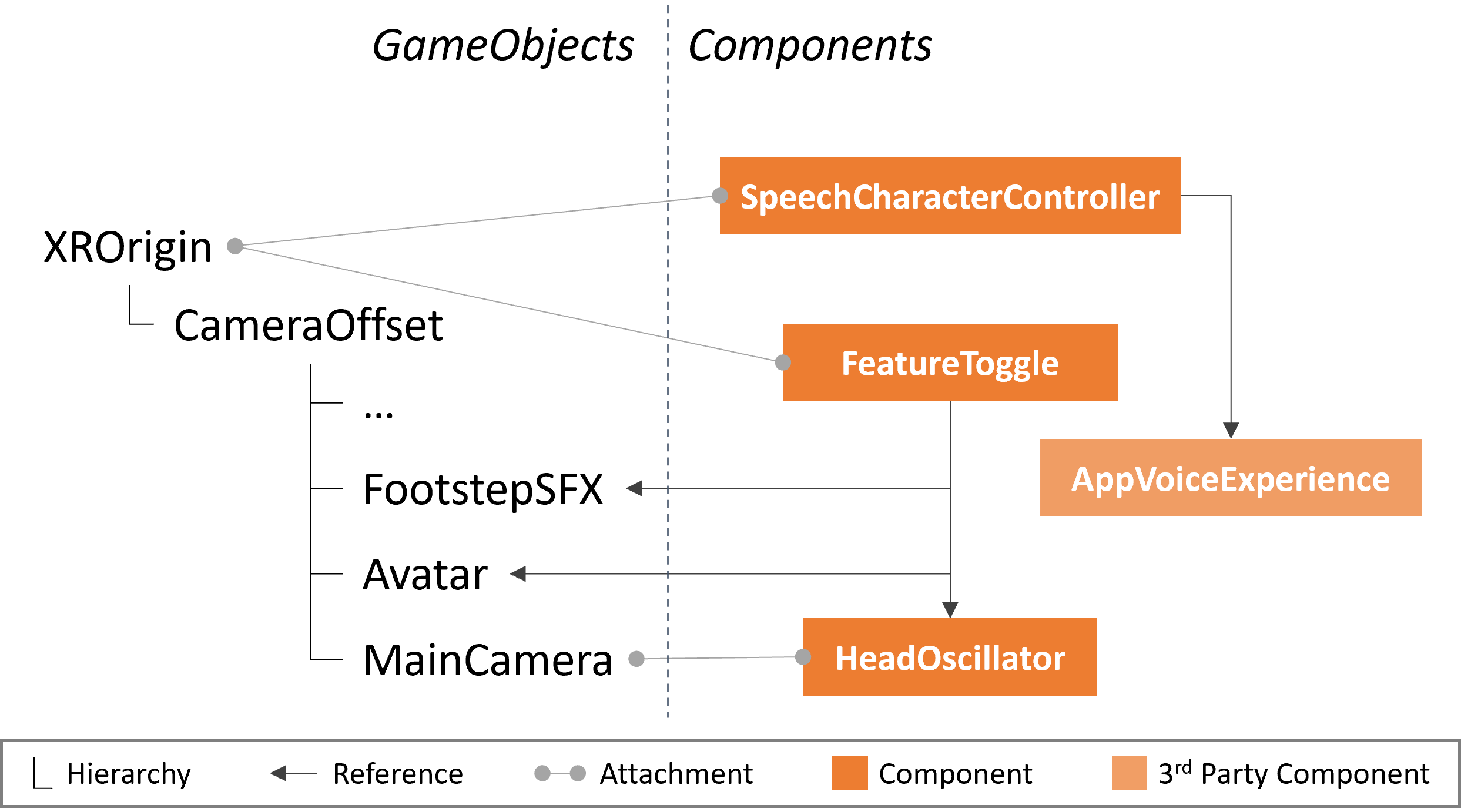}
    \caption{Unity-based Experimental System with Oculus Integration SDK and Wit.ai service. The left side displays the GameObject hierarchy, while the right side illustrates component attachment to GameObjects and their references.}
    \label{fig:architecture}
\end{figure}

The experimental system allowed users to move forward and backward as well as to turn left and right. Forward and backward movements were initially implemented as a sliding looking\cite{di2021locomotion} locomotion technique with continuous linear movement and, later, at Taheri's request, the system was updated to support three different speed settings for forward and backward movements. One keyword activated the movement forward, another activated the movement backward, and those remained active until the user said a third keyword to stop. The user could speed up and down by repeating the keyword for the current directions. The most comfortable linear speed was $1$ meter per second, allowing for 3 increment levels of $0.25$ meters per second. While slower than the average human walking speed~\cite{lai2020cognitive}, this speed value was used by Lécuyer et al.\cite{lecuyer2006camera} and found to be sufficient for Taheri to use voice inputs effectively to avoid obstacles and reach destinations at a satisfactory speed. The angular movement was initially implemented with a snap-turn technique of $45$ degrees around the user's vertical axis but was later replaced with a continuous turn of $45$ degrees per second from the current orientation to a $30$-degree displacement. Two keywords activated the turning, one for each direction. Furthermore, users could move their heads to turn in the direction they were looking at, allowing for fine adjustments of the turning direction. The HMD tracking only influenced the camera rotations and translations that had to be achieved using forward and backward linear movements activated through speech. The locomotion behavior is implemented by our \textit{SpeechCharacterController} component depicted in Fig. \ref{fig:architecture}.

The virtual environment in our system has a circular region of $100$ meters in diameter with four sections in the cardinal directions. Each of the sections has different conditions that make the user adapt their walking strategy. In the North direction, there is a soccer field with a ball so the user can kick it and try to score a goal. To enhance the user's ability to hit the ball, we increased its size to a radius of up to $90 cm$. This adjustment allowed Taheri to interact effectively with the ball, even with limited locomotion inputs. In the South, there is a sinuous tunnel where the user has to turn in a constrained space of $2.5$ meters wide to arrive at a secret chamber. In the East, there is a forest with sparse trees, so the user is stimulated to take turns to avoid the trees as they move. In the West, a hill with a windmill on top incentivizes the user to walk uphill and experience the view. The regions provide variability of walking conditions to the environment and encourage the user to walk around the virtual environment, but there was no goal or game associated with reaching any of the sections. The Unity assets used in the environment are freely available online\footnote{https://www.syntysearch.com/}\footnote{https://assetstore.unity.com/publishers/25353}.

\begin{figure}[!t]
    \centering
    \begin{tabular}{cc}
         
        \includegraphics[width=0.29\linewidth,alt={A person seated in a wheelchair, wearing a VR headset. The setting is an indoor space with multicolored carpeted flooring.}]{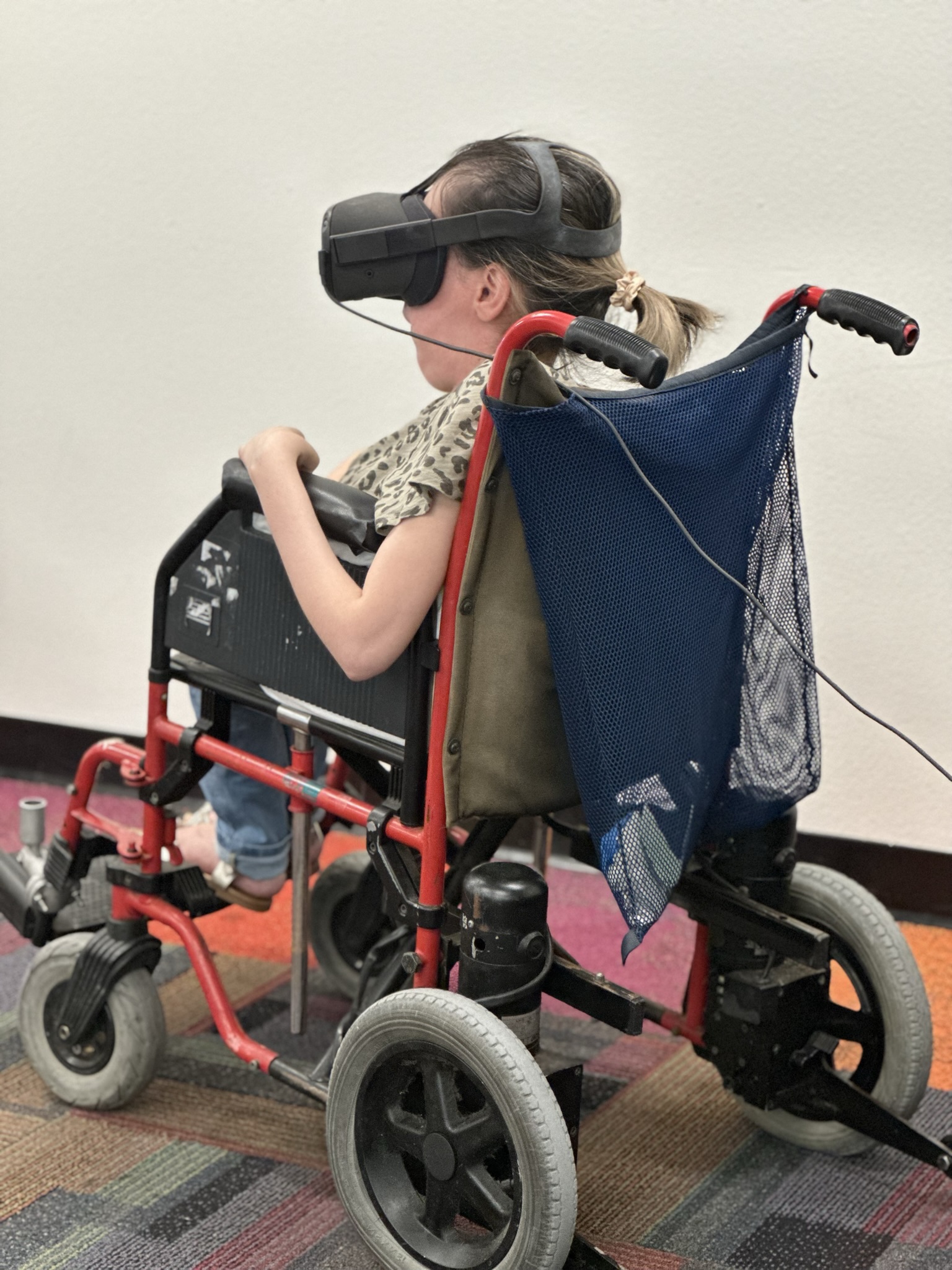} & \includegraphics[width=0.64\linewidth,alt={A 3D-rendered humanoid robot, reflected in a mirror within a simplistic virtual landscape with grassy field, trees, and hills in the background.}]{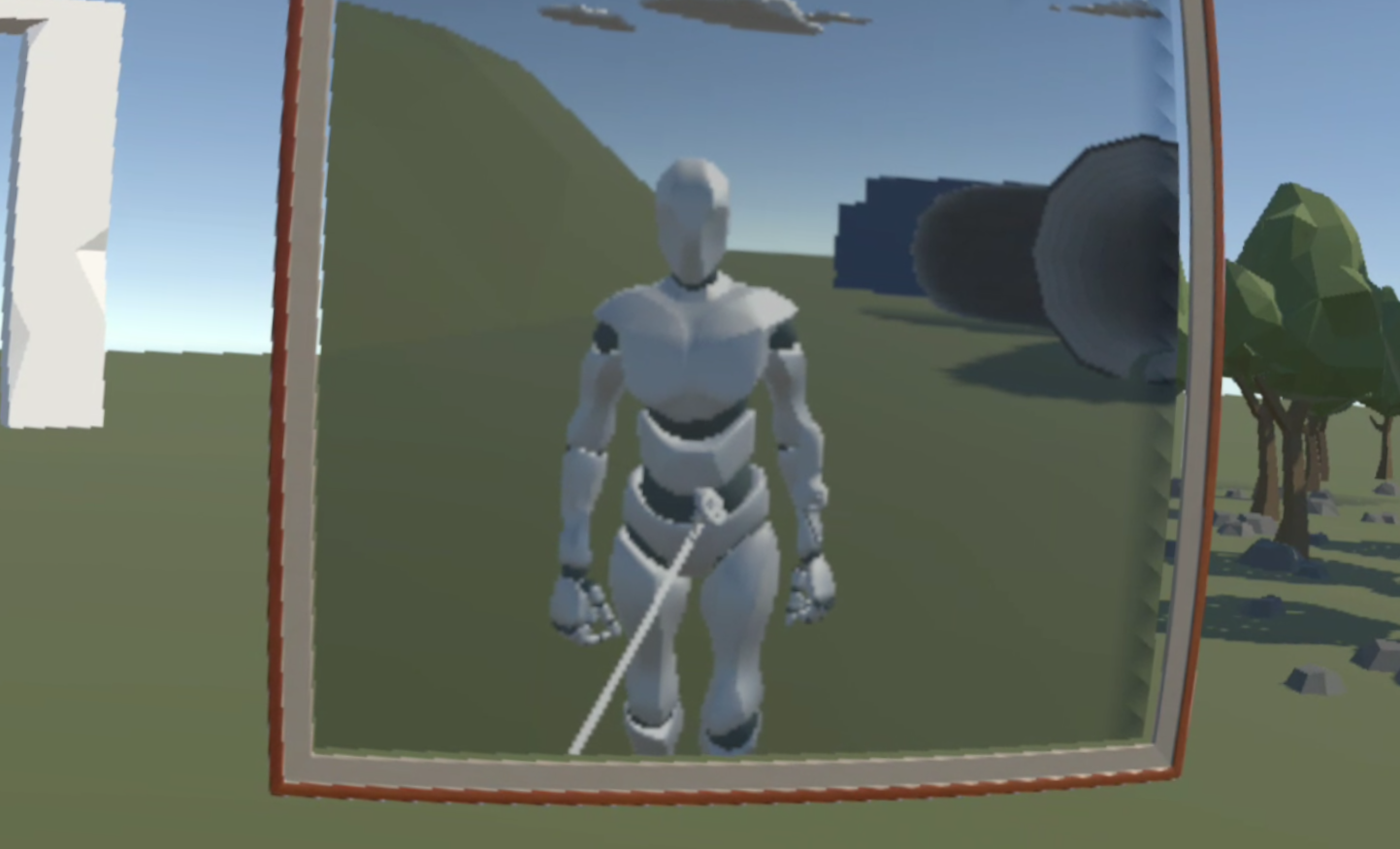}\\
        (a) & (b)
    \end{tabular}
    \caption{(a) Taheri testing the experimental system. (b) Virtual mirror in the experimental system.}
    \label{fig:Fey-mirror}
\end{figure}

Extremity joint tracking and inverse kinematics can enable full-body movement reconstruction in VR avatars and offer benefits in terms of embodiment and presence~\cite{caserman2019analysis}. However, this method was not practical in our scenario. Taheri faces motor restrictions that impede real-life walking, therefore tracking their body would be ineffective for simulating a walking experience. As an alternative to provide visual stimulus in the walking simulation, our system includes a full-body avatar with a walking animation, both freely available on  Adobe Mixamo\footnote{https://www.mixamo.com/}. The user can see the avatar in their peripheral vision and also through a virtual mirror (shown in Figure~\ref{fig:Fey-mirror}:b). The avatar's head was visible only through the mirror, being kept out of sight by adjusting the camera near-clip beyond the head. In future attempts to replicate this work, researchers may opt to cull the back faces of the avatar's head instead. We also adopted a head oscillation movement as additional visual stimuli. This was achieved by moving the camera around the initial local position to simulate gait-induced head oscillation, as proposed by Lecuyer et al.~\cite{lecuyer2006camera}. Lecuyer et al.'s head oscillation model comprises a three-dimensional amplitude (meters) vector parameter and a period determined by the ratio of stride length (meters) to advance speed (meters per second). The values for this parameter adopted during our iterations are documented in Table~\ref{tab:params}. The head oscillation mode preferred by Taheri had an amplitude of $5$ centimeters in all directions, stride length of $80$ centimeters, constant advance speed of $1$ meter per second, and a period of $1.6$ seconds. The head oscillation behavior is implemented by our \textit{HeadOscillator} component attached to the camera as shown in Fig.~\ref{fig:architecture}. As a complementary auditory stimulus, each footstep produces a sound that mimics stepping on grass, synchronized with the user's gait. Footstep sound effects were preferred when triggered in sync with all lateral oscillation extremes of the head, i.e., at every $0.8$ seconds. All of these stimuli can be toggled or tweaked to create different experimental conditions. The footstep sound effect is played by a \textit{FootstepSFX} GameObject, and the \textit{Avatar} GameObject holds the 3D model of the body. Both of them can be turned on and off by the user thanks to the \textit{FeatureToggle} component. The relationship of these components is illustrated in Fig.~\ref{fig:architecture}.

\subsection{Data Analysis}

Taheri and Caetano both contributed to the analysis of the data. Taheri, our single participant in this study, reported her perception and experience with the experimental system over the course of $9$ days in a diary. Across this period, a total of 2661 words were collected. The authors, including Taheri and Caetano, independently coded the diary entries and derived their own set of themes. Subsequently, they collaboratively refined and merged these themes into a unified set~\cite{braun2012thematic}. 

In addition to the identified themes, we also performed a text sentiment analysis across each iteration. We initially used an online tool~\footnote{https://huggingface.co/bhadresh-savani} based on DistilBERT~\cite{sanh2019distilbert}. The sentiment scores range from $0$ to $1$. A score near $0$ indicates a low likelihood, and near $1$ indicates a high likelihood that the text conveys this sentiment. In order to further understand the sentiment conveyed through Taheri's diary entries, we employed the Natural Language Toolkit (NLTK) package~\footnote{https://www.nltk.org/} in Python, which facilitated the deployment of two different sentiment analysis methodologies: VADER (Valence Aware Dictionary and sEntiment Reasoner)~\cite{hutto2014vader} and TextBlob~\cite{textblobTextBlobSimplified}. VADER, a lexicon, and rule-based sentiment analysis tool, was utilized to quantify sentiment in Taheri's entries, providing scores for positivity, negativity, and neutrality, along with a compound score. It requires no training data and is adept at analyzing text on social networks, among others. Positive, negative, and neutral scores are in the range of $[0, 1]$ for each aspect. Each score represents the proportion of the text that falls into these categories. Therefore, a higher positivity score indicates a higher proportion of the text is assessed as positive, and similarly for the negativity and neutrality scores. The compound score, on the other hand, is computed by summing the valence scores of each word in the lexicon, adjusted according to the rules, and then normalized to be between $-1$ (most extreme negative) and $+1$ (most extreme positive). Conversely, TextBlob, grounded on the Naive Bayes algorithm, delivered not only sentiment polarity scores but also subjectivity scores for each entry. The polarity score ranging from $-1$ (extremely negative) to $+1$ (extremely positive) served as an indicator of the sentiment orientation of the text. The subjectivity score, ranging from $0$ to $+1$, denotes the extent to which personal feelings or opinions are expressed in the text.

We also logged the combination of active features and head pose every second in a log file, enabling us to extract statistics and plot the trajectory of each trial.

To provide a clear overview of the iterative development process of our VR system, two tables have been included. Table~\ref{tab:changes-feedback} presents a summary of the modifications made in each iteration, along with Taheri's feedback. It offers insights into the evolving user experience and system refinement. Table~\ref{tab:params}, on the other hand, focuses on the specific parameter values adjusted across iterations, including voice commands, base walking speed, and head oscillation. This table gives a technical perspective on the changes, showcasing the incremental adjustments in system settings that were guided by Taheri's feedback.

\section{Results}

\subsection{Thematic Analysis}

\begin{table}[!t]
\begin{tabular}{|c|c|c|c|}
\hline
\textbf{Themes}                    & \textbf{Frequency} & \textbf{Ratio}   & \textbf{Variance} \\ \hline
Emotional Engagement      & 78        & 48.44\% & 31       \\ \hline
Mental Model of Walking   & 19        & 11.8\%  & 5.11     \\ \hline
User Interface \& Control & 10        & 6.21\%  & 0.86     \\ \hline
Embodiment \& Presence    & 20        & 12.42\% & 6.94     \\ \hline
Agency \& Control         & 28        & 17.39\% & 4.61     \\ \hline
Cybersickness             & 4         & 2.48\%  & 1.02     \\ \hline
\end{tabular}
\caption{Frequency and ratio of themes and their variance across iterations.}
\label{tab:freq-ratio-var}
\end{table}

Taheri's diary chronicling her experience using VR to walk for the first time provides a rich account full of insights. Several key themes emerged from a close reading of her diary. Table~\ref{tab:freq-ratio-var} illustrates the identified themes' frequency, ratio, and variance. This table provides a statistical overview of how frequently each theme was mentioned, its proportion relative to the total thematic mentions, and the variance indicating the fluctuation of each theme across different iterations.

\subsubsection{Emotional Engagement}
Throughout the experiment, Taheri reported diverse emotional responses, ranging from excitement to frustration. During the first session, a sense of ``excitement and nervousness'' dominated the experience, likely due to the novelty of the VR walking experience and the inherent expectations. As articulated at the beginning of her diary after Iteration 1, on Day 1:

\begin{quote}
    ``I felt a mix of excitement and nervousness since I didn't know what to expect, having never walked before in my life.''
\end{quote}

While initial excitement was high, technical challenges such as speech recognition errors led to some frustration. She remarked on the same day:

\begin{quote}
    ``My excitement waned a bit, but I was determined.''
\end{quote}

Taheri was ``determined'' to move forward despite these setbacks, illustrating a sense of resilience. 

The walking speed also had a significant impact on Taheri's emotional state. Initially, the ``painfully slow'' speed was a point of frustration. However, upon gaining control over the speed, Taheri felt ``freedom and agency,'' describing it as ``amazing to move however fast or slow I wanted on a whim.''

Emotional engagement is crucial for the user experience, especially for participants who have never walked before. The findings suggest that while the VR experience can elicit strong positive emotions, technical limitations can also lead to negative emotions like frustration, affecting the overall engagement level.

\subsubsection{Mental Model of Walking}
Even though Taheri had never physically experienced walking, she had her own mental model of what that experience would feel like. In the early iterations, she found the experience somewhat disorienting due to the lack of synchronization between visual and auditory elements. Specifically, on Day 3 and Iteration 3, she observed that the ``footsteps didn't quite sync up with the head bobbing,'' leading to a disrupted sense of realism. Taheri noted in her diary:

\begin{quote}
    ``The footsteps didn't quite sync up with the head bobbing. It was like hearing someone else's steps rather than my own.''
\end{quote}

The sound of the steps also played a crucial role in matching the virtual experience to Taheri's mental model. Initially, the footsteps were out of synchronization, creating a ``sense of disconnect.'' Over time, this was rectified, and Taheri noted that the ``perfected motions and footstep sounds'' enhanced her experience. The reported feeling of ``disconnect'' could be explained by a mismatch between the mental model and the sensory stimulus which is known to attenuate the sense of agency~\cite{wen2022sense, legaspi2019bayesian, moore2012sense}.

Head oscillation or bobbing was another feature that contributed to the sense of realism. Initially, Taheri found the head bobbing ``unusual'' but still ``amazing''  in Iteration 2, on Day 2. Over time, Taheri experimented with different settings and found that a moderate level of head bobbing felt ``comfortable'' and positively contributed to the experience. By Iteration 9, Day 9 she expressed:

\begin{quote}
    ``[..] at 1, I feel as if I'm truly walking, just like a person without a disability. It feels amazing and I love it.''
\end{quote}

The sense of realism in a VR walking experience is integral for immersion. Taheri's feedback highlights the need for better integration between auditory and visual elements to foster a more naturalistic experience.

\subsubsection{User Interface and Control}
Voice command recognition was a significant concern for Taheri, particularly in the early iterations. The system frequently confused commands, such as mistaking both ``walk'' and ``left'' for ``right,'' making the experience cumbersome and less intuitive. In Iteration 1 on Day 1, she expressed frustration:

\begin{quote}
    ``The system struggled to recognize my command to ‘walk’ forward, mistaking it for ‘right’ almost 99\% of the time.''
\end{quote}

This statement reflects the initial challenges encountered with voice command recognition, highlighting the need for refinement. Over subsequent iterations, all these issues were gradually addressed, leading to a more intuitive experience by Iteration 8, Day 8.

For physically disabled users who rely on voice commands, the accuracy and reliability of the interface are paramount. Failures in command recognition can significantly impair the user experience and lead to disengagement. 

\subsubsection{Embodiment and Presence}
The sense of embodiment evolved significantly across iterations, influenced by various factors. Notably, the introduction of a virtual avatar in Iteration 3 on Day 3 seemed to enhance Taheri's sense of embodiment. However, it introduced new distractions, such as the head appearing in the view due to the camera position. This issue was adjusted in Iteration 4, on Day 4, by increasing the camera near-clip beyond the head. Reflecting on the sense of embodiment produced by the Avatar in Iteration 3, on Day 3, Taheri reported:

\begin{quote}
``Looking down and seeing my legs moving as I ascended [the hill] was magical! This produced the strongest feeling yet of truly walking myself up the slope.'' 
\end{quote}

This reflection highlights the profound impact of the avatar on Taheri's sense of embodiment, enhancing the feeling of walking up a hill. Additionally, the virtual mirror allowed Taheri to see the reflected virtual movements of her virtual body, as expected of a physical mirror. Taheri recognized the reflected body as being her representation but not an accurate depiction of her physical body. In her diary, following Iteration 9, Day 9, the final iteration, Taheri articulated her feelings about this representation:

\begin{quote}
``I'm looking at my reflection - albeit an avatar robot - but it's still me in the mirror, [...]. It's a strange yet wonderful feeling to see myself standing and walking when the real me sits immobile.''
\end{quote}

The mirror might have induced cognitive dissonance by presenting an image that was at odds with the participant's real-world bodily experience. While it offered a novel and thrilling opportunity to ``stand,'' it also conflicted with her long-standing embodied experience of being in a wheelchair. This duality could have limited the sense of full embodiment within the VR environment.

The settings for head oscillation influenced Taheri's sense of embodiment. Allowing Taheri to adjust these settings enabled her to find the most comfortable configuration. When set to zero, the experience resembled ``rolling my wheelchair,'' but adjusting it to a moderate setting made Taheri feel as though she was ``truly walking''. The synchronization between the head oscillation and the footstep sound was also reported as impacting the sense of embodiment:

\begin{quote}
``The footsteps didn't quite sync up with the head bobbing. It was like hearing someone else's steps rather than my own. This mismatch disrupted what I imagined to be a natural rhythm of walking, creating a sense of disconnect.''
\end{quote}

This statement, from Iteration 3, Day 3, highlights the ongoing adjustments in the VR system and their impact on the sense of embodiment. The sense of embodiment and presence is crucial for VR applications, especially those aiming to replicate real-world experiences like walking. While the avatar increases the sense of presence, attention to detail is needed to avoid distractions that could break immersion.

\subsubsection{Agency and Control}
Taheri's sense of agency evolved significantly throughout the diary entries. During the initial iterations, as evidenced in Iteration 1, Day 1, the inability to accurately execute voice commands led to a limited sense of agency. Taheri felt confined by the system's inaccuracies, as evidenced by the struggles with the ``walk'', ``left'', and ``back'' commands.

As the study progressed, improvements in voice recognition and the introduction of speed controls significantly enhanced Taheri's sense of agency. The successful implementation of the ``go'' command for acceleration was a pivotal moment. Taheri expressed extreme happiness and a ``wonderful sense of freedom and agency,'' indicating a high level of control over her virtual experience. She specifically remarked upon in Iteration 8, Day 8:

\begin{quote}
    ``Using `go' for acceleration was the best choice. I was extremely happy to be able to control my walking pace. [...] It felt amazing to move however fast or slow I wanted on a whim.''
\end{quote}

She also had noted in Iteration 4, Day 4:
 
\begin{quote}
    ``For the first time in my life, I felt what it was like to climb a hill under my own power. I can't wait to explore more tomorrow!''
\end{quote}

A strong sense of agency is essential for a rewarding VR experience, especially for individuals who have physical disabilities. The ability to control one's speed or direction in the virtual environment can profoundly impact the user's emotional state and overall satisfaction with the VR experience. Therefore, optimizing the user interface for intuitive interactions is crucial for enhancing the user's sense of agency.

\subsubsection{Cybersickness}
Interestingly, despite the complexities involved in simulating walking for someone who has never physically experienced it, Taheri reported no instances of motion sickness. The first instance was during Iteration 2, Day 2, when she explicitly stated two times in her diary, once when she felt head oscillation to be ``exaggerated'':

\begin{quote}
    ``[head oscillation] oddly reminded me of a past experience of sitting on a horse with my cousin. Fortunately, there was no motion sickness at all.''
\end{quote}

And another time Iteration 7, Day 7 when she was trying continuous turning as she explicitly noted:

\begin{quote}
    ``I had been forewarned that this method of turning might induce motion sickness, but I did not experience any discomfort.''
\end{quote}

\subsection{Sentiment Analysis}
This subsection delves into the sentiment analysis results obtained from three different methodologies: DistilBERT, VADER, and TextBlob. The emotional landscape across different iterations of the study will be explored, highlighting specific emotions and events. Each method offers a unique perspective on Taheri's experience, contributing to a comprehensive understanding of emotional responses to virtually walking.

\subsubsection{Specific Emotions and Corresponding Events}
\begin{figure}[t!]
    \centering
    \includegraphics[width=1\linewidth, alt={A line graph shows fluctuations in seven sentiments across 9 iterations, with scores from 0 to 1. Each sentiment is depicted with a unique color and line style.}]{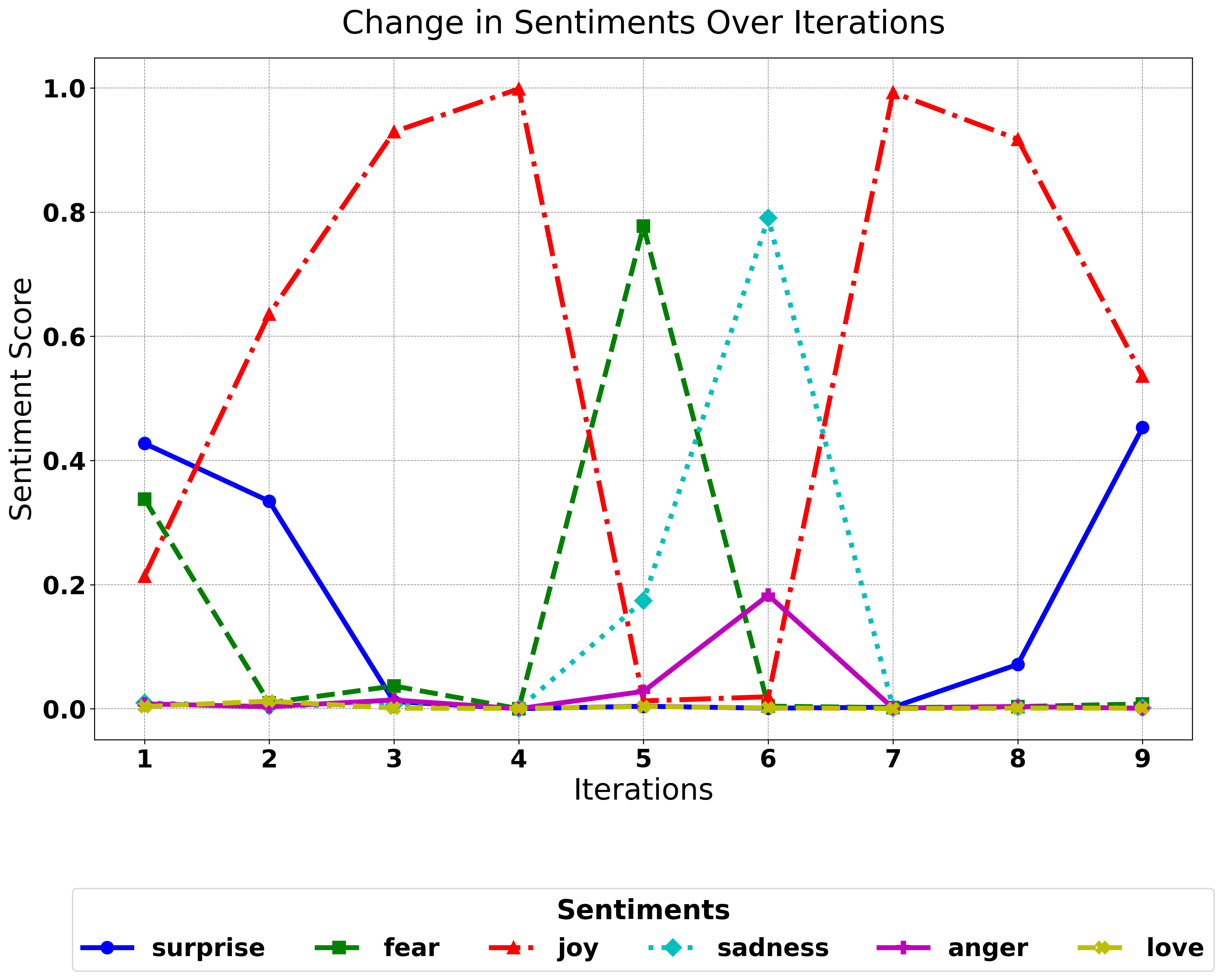}
    \caption{DistilBERT sentiment scores over $9$ iterations: variations in emotions—surprise, fear, joy, sadness, anger, and love—across iterations in the graph align closely with significant events and system enhancements, providing an understanding of Taheri's emotional journey.}
    \label{fig:DistilBERT-plot}
\end{figure}

As illustrated in Figure~\ref{fig:DistilBERT-plot}, the sentiment of \emph{``surprise''} was notably high during the first and last iterations. During the initial exposure to walking in VR, Taheri's diary revealed a mixture of excitement and nervousness, corroborating a high \emph{``surprise''} sentiment score of $0.43$. The concluding iteration led to another spike in \emph{``surprise''}  with a score of $0.45$, likely due to the integration of all previous settings, which culminated in a fulfilling and somewhat unexpected final experience. 

The emotion of \emph{``fear''} spiked significantly in the fifth iteration. This is aligned with Taheri's diary entry, which discussed a newfound concern about the monotony of rather long-distance virtual walks. Taheri directly related this feeling to the helplessness she often experiences while using a wheelchair and someone else is pushing it in the real world. This connection between virtual and real-world experience provides valuable context for the elevated \emph{``fear''} score of $0.78$.

The levels of \emph{ ``joy''} were remarkably elevated during the third and seventh iterations, with scores of $0.93$ and $0.99$, respectively. The third iteration introduced bug fixes and options to toggle various features, while the seventh incorporated continuous linear turning and speed controls. Both of these enhancements likely contributed to the high scores of \emph{``joy''} expressed in the graph.

Interestingly, the sixth iteration showed a complex emotional landscape with elevated levels of \emph{``sadness''} and \emph{``anger''}, with scores of $0.79$ and $0.18$, respectively. This iteration introduced speed controls, and diary entries hinted at possible difficulties or dissatisfaction with this new feature.

Throughout the study, the sentiment of \emph{``love''} remained low, most likely because this emotion was not a relevant factor in this context.

\subsubsection{VADER and TextBlob: A Detailed Emotional Journey}
\begin{figure}[t!]
    \centering
    \begin{tabular}{c}
        \includegraphics[width=\linewidth,alt={A line graph with iterations on the x-axis, from 1 to 9 and sentiment scores on the y-axis, from 0 to 1. It shows four sentiment trends: Positive (green line), Neutral (orange dashed), Negative (brown dotted), and Compound (purple dash-dot).}]{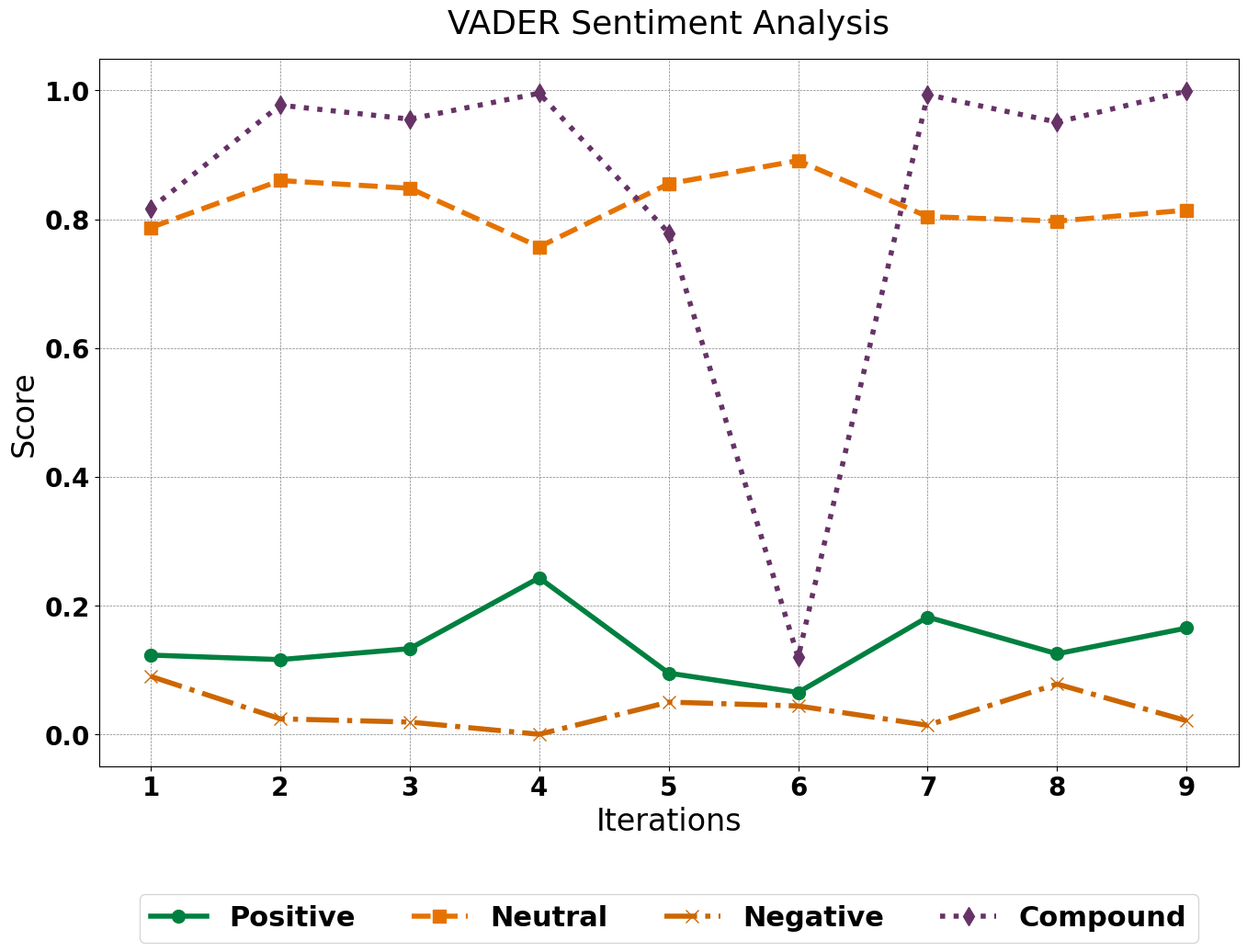} \\
        (a) VADER Sentiment Scores \\
        \\ 
        \includegraphics[width=\linewidth,alt={A line graph with two lines representing Polarity (blue) and Subjectivity (red dashed) over nine iterations. The x-axis shows iterations 1 through 9, and the y-axis shows scores from 0 to 1.}]{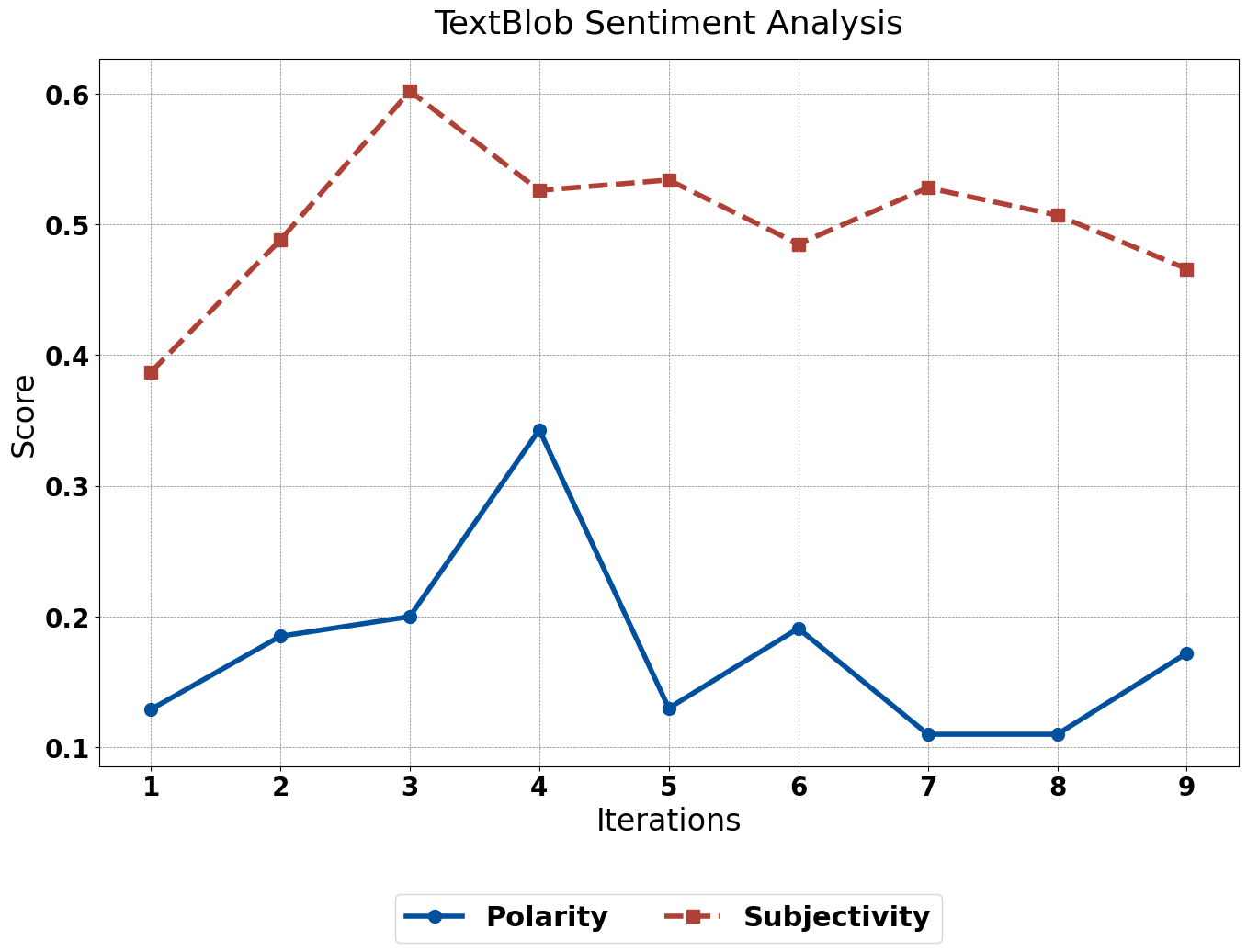} \\
        (b) TextBlob Sentiment Scores
    \end{tabular}
    \caption{Sentiment scores over $9$ iterations: (a) VADER sentiment scores showing the proportions of positive, neutral, and negative sentiments, as well as the compound score, across all iterations. (b) TextBlob sentiment scores showing polarity and subjectivity scores across all iterations.}
    \label{fig:VADER-TextBlob}
\end{figure}

\emph{Initial Encounters: Cautious Optimism (Iterations 1-3)}.\ Both VADER (Figure~\ref{fig:VADER-TextBlob}:a) and TextBlob (Figure~\ref{fig:VADER-TextBlob}:b) sentiment analyses aligned with the DistilBERT findings of a mix of excitement and apprehension during the initial iterations. VADER's compound scores ranged from 
$0.82$ to $0.95$, indicating a generally positive yet cautious sentiment. TextBlob's polarity echoed this trend, ranging from $0.13$ to $0.2$.

\emph{Emotional Highpoint: Awe and Freedom (Iteration 4)}.\ In the fourth iteration, VADER's positive score soared to $0.24$, and the compound score reached $0.99$, corroborating the heightened sentiment of ``joy'' reported by DistilBERT. TextBlob's polarity peaked at $0.34$, confirming a highly positive emotional state. These peaks are visible in Figure~\ref{fig:VADER-TextBlob}:(a and b).

\emph{Emotional Complexities: Mixed Feelings (Iterations 5-6)}.\ The fifth and sixth iterations saw a decline in overall positive sentiment, which is consistent with the spike in ``fear'' observed in DistilBERT. VADER's positive scores dropped to $0.09$ and $0.06$, respectively, while the compound scores also decreased. TextBlob's polarity scores mirrored this trend, as shown in Figure\ref{fig:VADER-TextBlob}:(a and b).

\emph{Final Iterations: Satisfaction and Reflection (Iterations 7-9)}.\ The final iterations witnessed a resurgence in positive sentiment, consistent with the elevated levels of ``joy'' in DistilBERT. VADER's positive scores rebounded to 
$0.18$ in the seventh iteration and remained relatively high through the ninth. The compound scores also increased, culminating in a near-maximum score of $0.999$ in the final iteration. TextBlob's polarity scores were consistent with this trend.

\emph{Subjectivity Across Iterations}.\ TextBlob's subjectivity scores provided additional depth to our understanding of Taheri's experience. The scores ranged from 
$0.39$ to $0.6$, indicating varying levels of personal engagement and emotional investment across iterations. These are well-captured in Figure~\ref{fig:VADER-TextBlob}:(a and b) for VADER and TextBlob, respectively.

\subsection{Trajectory Analysis}

\begin{figure}[t!]
    \centering
    \includegraphics[width=\linewidth, alt={A virtual green landskape with a dotted trajectory line passing through labeled landmarks like a windmill, soccer field, tunnel, mirror, and forest, with starting and ending points marked.}]{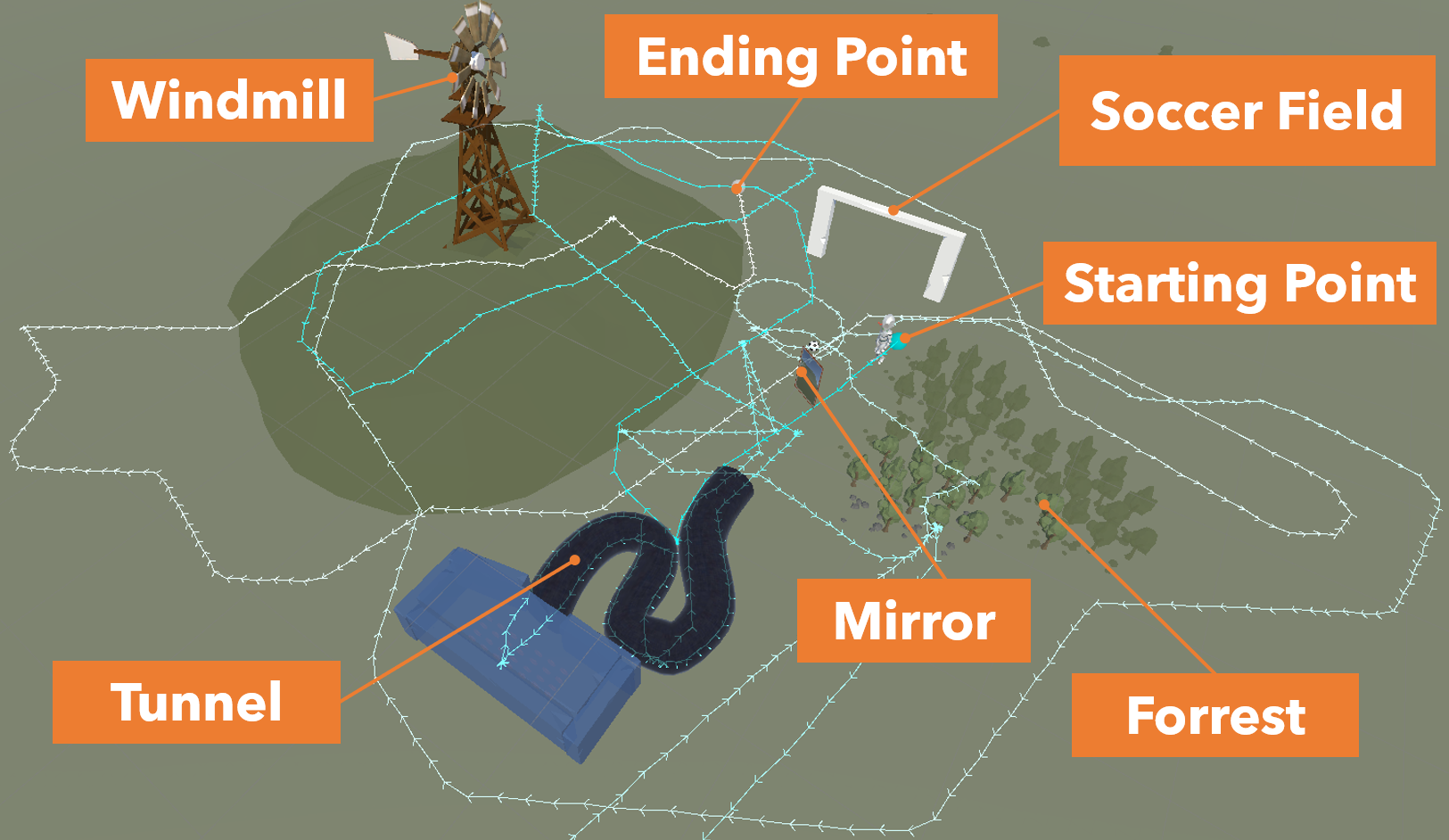}
    \caption{The virtual environment provided the setting for the virtual walking experience, featuring landmarks and props that motivated Taheri to explore. Her trajectory began at the cyan sphere and ended at the white sphere, with directional arrows indicating the path's orientation.}
    \label{fig:trajectory}
\end{figure}

Taheri tested the experimental system for a total of 3 hours and 45 minutes, covering a distance of 5.725 kilometers during the span of 9 days. On average, iterations lasted 12 minutes, with the longest iteration lasting 26 minutes. The average iteration length was 370 meters, while the longest iteration extended to 1.18 kilometers. The average speed during the iterations was approximately 0.5 meters per second. These are statistics from the third iteration onward; the logging feature was introduced in that iteration. Figure~\ref{fig:trajectory} shows the trajectory Taheri walked in the virtual environment.

\section{Discussion}
This exploratory study provides insights into the subjective experience of virtually walking for individuals who have never walked before. Our participatory design approach enabled an iterative process where the feedback from one of the researchers in our team, who was the study participant and a member of the target user group, directly shaped system enhancements to better meet her needs and desires. The themes that emerged from the diary study analysis highlight key considerations for creating compelling and emotionally engaging VR walking experiences for people with disabilities.

The choice of the diary method for data analysis played a crucial role in this research. This approach allowed Taheri to document thoughts and feelings in realtime, providing a rich and nuanced understanding of her experience with the VR system. Unlike observation or video analysis, the diary method empowers the participant to express her perspective in her own words, offering insights that might not be evident through external observation. Additionally, compared to post-trial interviews, diaries can capture immediate reactions and changes in perception across different stages of the experiment, which is crucial for understanding the evolving experience of a lifelong wheelchair user in a VR walking simulation. This method's introspective and longitudinal nature made it particularly suitable for this exploratory study, where the participant's subjective experience was paramount~\cite{bolger2003diary, elliott1997use}.

The sentiment analysis reinforces the importance of managing user expectations and technical limitations to avoid frustration. As observed in the initial iterations, inaccuracies in speech recognition and lack of control over speed disrupted immersion and caused negative reactions. However, once these issues were addressed, emotional engagement improved dramatically. The final iterations produced high levels of ``joy'', indicating satisfaction with the experience. Our findings reveal an interplay between the senses of agency, embodiment, and presence in simulating the experience of walking. Taheri's mental model of walking had to be carefully considered in syncing visual and auditory elements like footsteps and head bobbing. A user able to walk in the physical space may also express discomfort when experiencing a misalignment between their intentions and sensorimotor stimuli, as indicated by models of the sense of agency~\cite{wen2022sense, legaspi2019bayesian, moore2012sense}. We speculate that in Taheri's case, her sense of agency predominantly stems from her mental model of walking, at the cognitive level, given her sensorimotor system's limitations. This underscores the importance of developing VR experiences that are congruent with the mental models of users facing mobility impairments.

Customization was key; options to tweak head oscillation and walking speed enhanced the sense of control and embodiment. The virtual body avatar increased embodiment but could also be distracting if not properly calibrated. Overall, a high degree of realism and synchronicity is needed to avoid breaking immersion. The life-long experience of disability colors the subjective experience of virtual walking. Connections were drawn between feelings of helplessness in the real world and virtually walking long distances. Managing expectations and emotional support may be as important as technical accuracy.

The introduction of the virtual mirror produced a unique reaction, allowing Taheri to see herself standing and walking despite her physical condition. 
She reported a ``strange yet wonderful feeling'' when looking at the virtual mirror. We attribute this perception to the previously demonstrated impact of tool embodiment on extrapersonal space estimations~\cite{scandola2019embodying}. We speculate that the virtual mirror allowed Taheri to estimate her extrapersonal space, but the avatar not depicting the wheelchair confused her sense of embodiment, producing the reported strange feeling. The sensation of ``strangeness'' could be attributed to the dissonance between Taheri's lifelong physical experience and the virtual representation of her body. For someone who has never walked, seeing oneself standing and walking in a mirror can be an unfamiliar experience, challenging long-held perceptions of self and mobility. The ``wonderful'' aspect likely stems from the empowering experience of virtually performing a physically unattainable action. This duality of emotions reflects the complex interplay between physical reality and virtual embodiment.

A critical consideration of this study is the concept of cybersickness which occurs due to decoupling virtual input control from motion. In the physical world, this decoupling for Taheri may manifest itself as the use of a joystick to move the wheelchair or having someone else push the wheelchair. In the virtual environment, a different form of decoupling occurs. By adopting speech as an input modality, there was a chance that the user's mental model of walking would be challenged by the decoupling of locomotion through a virtual input. Interestingly, despite our anticipation of potential cybersickness, Taheri reported no such symptoms. Although the reasons for cybersickness are debated~\cite{aykent2011study}, we believe that this was related to a comfortable speed parameterization, perhaps because the levels of sensory mismatch in the VR experience were similar to those experienced by Taheri in the wheelchair.

\section{Limitations and Future Work}
Despite novel findings, this study is not without limitations. First, the results are derived from a single participant, which, while providing depth, may not capture the breadth of experiences of all individuals with similar situations, i.e., not having prior experience of walking in real life. Second, while the study took into account several adjustable parameters in the VR environment, there are potentially other elements, not explored in this research, that could influence a user's experience (e.g., haptic feedback). While the generalizability of our results to healthy users requires additional study, we expect our findings to be a starting point for creating VR experiences that align with the mental models of 
individuals with congenital mobility impairments.

Our experimental design included a single participant, a co-author who is part of the target group that has never experienced real-world walking.
We acknowledge the benefits and limitations of membership and disclosure in HCI research with marginalized communities~\cite{liang2021embracing} and encourage readers to interpret our findings carefully.

\section{Conclusion}
This study contributes to our understanding of designing VR experiences for individuals with congenital mobility impairments. Through an iterative participatory design approach and rich data from diary entries, we found that the user experience is influenced by factors ranging from technical details to emotional responses. The use of a humanoid avatar, head oscillations, and synchronized footsteps, as visual and auditory stimuli matched the mental model of the participant. These findings have implications for HCI designers and researchers aiming to create more inclusive and emotionally engaging VR systems.

\bibliographystyle{abbrv-doi}

\bibliography{arxiv}
\end{document}